\documentstyle[a4,12pt]{article}
\begin{document}


\vspace{0.2cm}

\begin{center}
{\large\bf Testing $CPT$ Invariance in $B^0_d$-$ \bar{B}^0_d$ and
$B^0_s$-$ \bar{B}^0_s$ Oscillations}
\end{center}

\vspace{0.2cm}

\begin{center}
{\bf Ping Ren} ~ and ~ {\bf Zhi-zhong Xing} \footnote{
E-mail: xingzz@ihep.ac.cn}\\
{\sl Institute of High Energy Physics, Chinese Academy of
Sciences, \\
P.O. Box 918, Beijing 100049, China}
\end{center}

\vspace{2cm}

\begin{abstract}
Recent CDF and D0 measurements of $B^0_s$-$\bar{B}^0_s$ mixing
make it possible to search for $CP$ violation and test $CPT$
symmetry in a variety of $B^{}_s$ decays. Considering both
coherent $B^0_d\bar{B}^0_d$ decays at the $\Upsilon (4S)$
resonance and coherent $B^0_s\bar{B}^0_s$ decays at the $\Upsilon
(5S)$ resonance, we formulate their time-dependent and
time-integrated rates by postulating small $CPT$ violation in
$B^0_d$-$ \bar{B}^0_d$ and $B^0_s$-$ \bar{B}^0_s$ oscillations. We
show that the opposite-sign dilepton events from either $C$-odd or
$C$-even $B^0_q\bar{B}^0_q$ states (for $q= d$ or $s$) can be used
to determine or constrain the $CPT$-violating parameter at a
super-$B$ factory. The possibility of distinguishing between the
effect of $CPT$ violation and that of $\Delta B = -\Delta Q$
transitions is also discussed.
\end{abstract}


\newpage

\section{Introduction}

A correlated $P^0 \bar{P}^0$ system, where $P$ may be either $K$,
$D$, $B^{}_d$ or $B^{}_s$, has been of great interest for the
study of $CP$, $T$ and $CPT$ symmetries in particle physics. In
the $K^0$-$\bar{K}^0$ mixing system, for instance, both indirect
$CP$ violation \cite{K1} and direct $CP$ violation \cite{K2} have
unambiguously been observed; the evidence for $T$ violation
\cite{K3} has been achieved; and $CPT$ invariance has been tested
to an impressive degree of accuracy \cite{PDG06}. The $\Delta Q =
\Delta S$ rule has also been examined in the semileptonic $K^0$
and $\bar{K}^0$ transitions. Beyond the $K^0\bar{K}^0$ system,
both indirect and direct signals of $CP$ violation have been
observed in a number of neutral $B^{}_d$ decays \cite{PDG06}; and
possible $CPT$ violation in $B^0_d$-$\bar{B}^0_d$ mixing has been
searched for at the KEK and SLAC $B$-meson factories \cite{BB}.
Although the phenomena of $CP$ violation have not been seen in the
$B^0_s$-$\bar{B}^0_s$ and $D^0$-$\bar{D}^0$ mixing systems, they
may show up and even surprise us in the near future at the LHC-$B$
\cite{Nakada}, $\tau$-charm \cite{Asner} and super-$B$
\cite{Yamauchi} factories.

The CDF \cite{CDF} and D0 \cite{D0} Collaborations have recently
reported their measurements of $B^0_s$-$\bar{B}^0_s$ mixing (both
the mass and width differences between the light and heavy
$B^{}_s$ mass eigenstates) at the Fermilab Tevatron Collider:
\begin{eqnarray}
\Delta M^{}_s & = & 17.77 \pm 0.10 ({\rm stat}) \pm 0.07 ({\rm
syst}) ~ {\rm ps}^{-1} \; ({\rm CDF} ~\cite{CDF}) \; , \nonumber
\\
\Delta \Gamma^{}_s & = & 0.13 \pm 0.09 ~ {\rm ps}^{-1} \; ({\rm
D0} ~\cite{D0}) \; .
\end{eqnarray}
This remarkable progress in experimental $B$ physics makes it
possible to search for $CP$ violation and test $CPT$ symmetry in
the $B^0_s\bar{B}^0_s$ system. As the magnitude of
$B^0_s$-$\bar{B}^0_s$ mixing is much larger than that of
$B^0_d$-$\bar{B}^0_d$ mixing \cite{PDG06}, its impact on the
time-dependent and time-integrated rates of $B^0_s$ and
$\bar{B}^0_s$ decays deserves attention. In particular, the $CP$-
and $CPT$-violating signals might get enhanced or suppressed due
to large $B^0_s$-$\bar{B}^0_s$ mixing. At a super-$B$ factory with
the luminosity ${\cal L} \sim {\rm a ~ few} \times 10^{36}{\rm
cm}^{-2}{\rm s}^{-1}$ \cite{Yamauchi}, both coherent
$B^0_d\bar{B}^0_d$ decays at the $\Upsilon (4S)$ resonance and
coherent $B^0_s\bar{B}^0_s$ decays at the $\Upsilon (5S)$
resonance will be studied down to the last detail.

The main purpose of this work is to formulate the decay rates of a
correlated $B^0_q\bar{B}^0_q$ state (either $q=d$ or $q=s$) in the
assumption that there are both small $CPT$ violation in
$B^0_q$-$\bar{B}^0_q$ oscillation and small $\Delta B = \Delta Q$
violation in semileptonic $B^0_q$ and $\bar{B}^0_q$ decays. Note
that possible effects of $CPT$ violation in neutral $B^{}_d$ and
$D$ decays have been analyzed in Refs. \cite{CPTB,Xing99,Grimus}
and Refs. \cite{CPTD,Xing97}, respectively; and possible effects
of $\Delta B = -\Delta Q$ and $\Delta C = -\Delta Q$ transitions
have been discussed in Ref. \cite{Sarma} and Ref. \cite{QD},
respectively. The present paper is different from the previous
ones not only because we are dealing with a new heavy
meson-antimeson mixing system (i.e., the $B^0_s\bar{B}^0_s$
system) but also because our discussions are essentially new in
two aspects. (1) We shall take into account both $C$-odd and
$C$-even $B^0_q\bar{B}^0_q$ pairs with $C$ being the
charge-conjugation parity of this coherent system, and calculate
both time-dependent and time-integrated rates of
$(B^0_q\bar{B}^0_q)^{}_C$ decays by assuming slight $CPT$
violation and small $\Delta B = -\Delta Q$ effects. Our analytical
results are more general and more useful than those obtained in
Refs. \cite{CPTB}--\cite{Sarma}. (2) We shall concentrate on the
opposite-sign dilepton asymmetries of $(B^0_q\bar{B}^0_q)^{}_C$
decays to investigate possible effects of $CPT$ violation and
$\Delta B = -\Delta Q$ transitions. It is worth remarking that an
opposite-sign dilepton event may be either $l^+_1 l^-_2$ (for
$l^{}_1 \neq l^{}_2$) or $l^+l^-$. Although the opposite-sign
dilepton events of $(B^0_d\bar{B}^0_d)^{}_C$ decays have been
considered in Refs. \cite{Xing99,Grimus}, our results for the
$B^0_s$-$\bar{B}^0_s$ mixing system have their own features and
implications due to the small $CP$-violating phase of
$B^0_s$-$\bar{B}^0_s$ mixing and the large values of $\Delta
M^{}_s$ and $\Delta \Gamma^{}_s$.

So far some interest has been paid to the possibilities of
exploring $CP$ violation and probing new physics in weak $B^{}_s$
decays at the $\Upsilon (5S)$ resonance
\cite{Xing98}--\cite{Xing00}, although the experimental
feasibility remains an open question. We expect that the future
super-$B$ factory can run at this interesting energy threshold
\cite{5S}. Then it will be possible to test $CPT$ invariance in
both $B^0_d$-$\bar{B}^0_d$ and $B^0_s$-$\bar{B}^0_s$ oscillations
by studying the coherent $B^0_d\bar{B}^0_d$ and $B^0_s\bar{B}^0_s$
decays.

\section{$CPT$ violation in coherent $(B^0_q\bar{B}^0_q)^{}_C$ decays}

The mixing or oscillation between $B^0_q$ and $\bar{B}^0_q$ mesons
can naturally arise from their common coupling to a subset of real
and virtual intermediate states. Hence the mass eigenstates
$|B^{}_{\rm L}\rangle$ and $|B^{}_{\rm H}\rangle$, where ``L"
(``H") denotes ``light" (``heavy"), are different from the flavor
(weak interaction) eigenstates $|B^0_q\rangle$ and
$|\bar{B}^0_q\rangle$. Taking account of both $CP$- and
$CPT$-violating effects in $B^0_q$-$\bar{B}^0_q$ mixing, one may
parametrize the correlation between $\{|B^{}_{\rm L}\rangle,
|B^{}_{\rm H}\rangle\}$ and $\{|B^0_q\rangle,
|\bar{B}^0_q\rangle\}$ states as follows:
\begin{eqnarray}
|B^{}_{\rm L}\rangle & = & \cos\frac{\theta}{2} ~
e^{-i\frac{\phi}{2}} |B^0_q\rangle + \sin\frac{\theta}{2} ~ e^{+
i\frac{\phi}{2}}
|\bar{B}^0_q\rangle \; , \nonumber \\
|B^{}_{\rm H}\rangle & = & \sin\frac{\theta}{2} ~
e^{-i\frac{\phi}{2}} |B^0_q\rangle - \cos\frac{\theta}{2} ~ e^{+
i\frac{\phi}{2}} |\bar{B}^0_q\rangle \; ,
\end{eqnarray}
where $\theta$ and $\phi$ are in general complex. For simplicity,
the normalization factors of $|B^{}_{\rm L}\rangle$ and
$|B^{}_{\rm H}\rangle$ in Eq. (2) have been omitted. $CPT$
invariance requires $\cos\theta =0$ or equivalently $\theta =
\pi/2$, while $CP$ conservation requires both $\theta =\pi/2$ and
$\phi =0$ \cite{Lee}
\footnote{As $CPT$ violation may simultaneously imply the
violation of Lorentz covariance in a quantum field theory
\cite{Greenberg}, the dependence of $\theta$ on the sidereal time
should in general be taken into account \cite{Kos}. For
simplicity, here we take $\theta$ as a constant by assuming that
the boost parameters of $B^0_q$ and $\bar{B}^0_q$ mesons are small
and the corresponding Lorentz violation is rotationally invariant
in the laboratory frame. In this approximation, our results are
valid as averages over the sidereal time, such that the effect of
Lorentz violation due to the direction of motion can be neglected.
A complete analysis, which requires incorporating
Lorentz-violating parameters directly into the phenomenology to
account for possible $CPT$ violation \cite{Kos2}, is beyond the
scope of this paper and will be done elsewhere.}.
The proper-time evolution of an initially pure $|B^0_q\rangle$ or
$|\bar{B}^0_q\rangle$ state is given by \cite{CPTB}
\begin{eqnarray}
|B^0_q(t)\rangle & = & e^{- \left (i M +\frac{\Gamma}{2} \right )
t} \left [g^{}_+(t) |B^0_q\rangle + \tilde{g}^{}_+(t)
|\bar{B}^0_q\rangle \right ] \; ,
\nonumber \\
|\bar{B}^0_q(t)\rangle & = & e^{- \left (i M +\frac{\Gamma}{2}
\right ) t} \left [g^{}_-(t) |\bar{B}^0_q\rangle +
\tilde{g}^{}_-(t) |B^0_q\rangle \right ] \; ,
\end{eqnarray}
where
\begin{eqnarray}
g^{}_{\pm}(t) & = & \cosh \left (\frac{ix^{}_q -y^{}_q}{2} \Gamma
t \right ) \pm \cos\theta \sinh \left (\frac{ix^{}_q -y^{}_q}{2}
\Gamma t \right ) \; , \nonumber \\
\tilde{g}^{}_{\pm}(t) & = & \sin\theta ~ e^{\pm i\phi} \sinh \left
(\frac{ix^{}_q -y^{}_q}{2} \Gamma t \right ) \; .
\end{eqnarray}
The definitions in Eqs. (3) and (4) are $M \equiv (M^{}_{\rm L} +
M^{}_{\rm H})/2$, $\Gamma \equiv (\Gamma^{}_{\rm L} +
\Gamma^{}_{\rm H})/2$, $x^{}_q \equiv \Delta M^{}_q/ \Gamma$ with
$\Delta M^{}_q \equiv M^{}_{\rm H} - M^{}_{\rm L}$, and $y^{}_q
\equiv \Delta \Gamma^{}_q/(2\Gamma)$ with $\Delta \Gamma^{}_q
\equiv \Gamma^{}_{\rm L} - \Gamma^{}_{\rm H}$, where $M^{}_{\rm
L,H}$ ($\Gamma^{}_{\rm L,H}$) denotes the mass (width) of
$B^{}_{\rm L,H}$. Taking account of $1/\Gamma \approx 1.52 ~{\rm
ps}$ \cite{D0}, we approximately obtain $x^{}_s \approx 27$ and
$y^{}_s \approx 0.1$ from the central values of $\Delta M^{}_s$
and $\Delta \Gamma^{}_s$ given in Eq. (1). In contrast, $x^{}_d
\approx 0.78$ \cite{PDG06} has been known but $y^{}_d$ has not
been measured for the $B^0_d$-$\bar{B}^0_d$ mixing system. We
stress that the experimental values of $x^{}_d$, $x^{}_s$ and
$y^{}_s$ are in good agreement with the standard-model predictions
\cite{Lenz}
\footnote{Note that $y^{}_d \approx 0.002$ and $y^{}_s \approx
0.06 \cdots 0.08$ are the updated predictions \cite{Lenz}. The
theoretical expectation $x^{}_s/x^{}_d \sim y^{}_s/y^{}_d \sim 35$
has partly been confirmed by current experimental data.},
although the uncertainty associated with $y^{}_s$ remains quite
large.

In order to calculate the proper-time distribution of coherent
$(B^0_q\bar{B}^0_q)^{}_C$ decays, we neglect the tiny final-state
electromagnetic interactions and assume $CPT$ invariance in the
direct transition amplitudes of semileptonic or nonleptonic
$B^0_q$ and $\bar{B}^0_q$ decays. Such an assumption can be
examined, without the mixing-induced complexity, by detecting the
charge asymmetry of semileptonic $B^{\pm}$ decays \cite{Xing99}.
We shall take into account possible $\Delta B =-\Delta Q$
transitions in our calculations. The latter can be described by
using a small parameter $\sigma^{}_l$ for a given semileptonic
decay mode,
\begin{eqnarray}
\langle l^+X^-_l|{\cal H}|B^0_q\rangle \; \equiv \; A_l \; , &&
~~~~~~~ \langle l^+X^-_l|{\cal H}|\bar{B}^0_q\rangle \; =\;
\sigma^{}_l
A_l \; ; \nonumber \\
\langle l^-X^+_l|{\cal H}|\bar{B}^0_q\rangle \; \equiv \; A^*_l \;
, && ~~~~~~~ \langle l^-X^+_l|{\cal H}|B^0_q\rangle \; =\;
\sigma^*_l A^*_l \; ,
\end{eqnarray}
where $\sigma^{}_l$ measures the $\Delta B =-\Delta Q$ effect and
$|\sigma^{}_l|\ll 1$ is expected to hold. $|\sigma^{}_l| \neq 0$
implies that it is in practice impossible to have a {\it pure}
tagging of the $B^0_q$ or $\bar{B}^0_q$ state through its
semileptonic decay (to $l^+X^-_l$ or $l^-X^+_l$). In general, the
amplitudes of $B^0_q$ and $\bar{B}^0_q$ decays into the
final-state $f^{}_i$ (either semileptonic or nonleptonic) are
denoted as $A^{}_{f^{}_i} \equiv \langle f^{}_i|{\cal
H}|B^0_q\rangle$ and $\bar{A}^{}_{f^{}_i} \equiv \langle
f^{}_i|{\cal H}|\bar{B}^0_q\rangle$. When the coherent
$(B^0_q\bar{B}^0_q)^{}_C \rightarrow f^{}_1 f^{}_2$ decays are
concerned, the following combinations
\begin{equation}
\xi^{}_C \; = \; e^{-i\phi} + C e^{i\phi}
\frac{\bar{A}^{}_{f^{}_1} \bar{A}^{}_{f^{}_2}}{A^{}_{f^{}_1}
A^{}_{f^{}_2}} \; , ~~~~~~~ \zeta^{}_C \; =\;
\frac{\bar{A}^{}_{f^{}_2}}{A^{}_{f^{}_2}} + C
\frac{\bar{A}^{}_{f^{}_1}}{A^{}_{f^{}_1}} \; ,
\end{equation}
where $C= \pm 1$, will be frequently used.

Now let us consider a correlated $B^0_q \bar{B}^0_q$ state at
rest. Its time-dependent wave function can be written as
\begin{equation}
\frac{1}{\sqrt{2}} \left [ |B^0_q ({\bf K}, t)\rangle |\bar{B}^0_q
(-{\bf K}, t)\rangle + C |B^0_q (-{\bf K}, t)\rangle |\bar{B}^0_q
({\bf K}, t)\rangle \right ] \; ,
\end{equation}
where $\bf K$ is the three-momentum vector of $B^0_q$ and
$\bar{B}^0_q$, and $C =\pm 1$ is the charge-conjugation parity of
this coherent system. If one of the two $B^{}_q$ mesons (with
momentum $\bf K$) decays to a final state $f^{}_1$ at proper time
$t^{}_1$ and the other (with $-{\bf K}$) to $f^{}_2$ at $t^{}_2$,
the amplitude of their joint decays is given by
\begin{eqnarray}
A \left (f^{}_1, t^{}_1; f^{}_2, t^{}_2 \right )^{}_C & = &
\frac{1}{\sqrt{2}} ~ e^{-\left ( i M + \frac{\Gamma}{2} \right )
\left (t^{}_1 + t^{}_2 \right )} \left \{ A^{}_{f^{}_1}
A^{}_{f^{}_2} \left [ g^{}_+(t^{}_1) \tilde{g}^{}_-(t^{}_2) + C
\tilde{g}^{}_-(t^{}_1)
g^{}_+(t^{}_2) \right ] \right . \nonumber \\
&& ~~~~~~~~~~~~~~~~~~~~~~~~ + A^{}_{f^{}_1} \bar{A}^{}_{f^{}_2}
\left [ g^{}_+(t^{}_1) g^{}_-(t^{}_2) + C \tilde{g}^{}_-(t^{}_1)
\tilde{g}^{}_+(t^{}_2) \right ] \nonumber \\
&& ~~~~~~~~~~~~~~~~~~~~~~~~ + \bar{A}^{}_{f^{}_1} A^{}_{f^{}_2}
\left [ \tilde{g}^{}_+(t^{}_1) \tilde{g}^{}_-(t^{}_2) + C
g^{}_-(t^{}_1)
g^{}_+(t^{}_2) \right ] \nonumber \\
&& \left . ~~~~~~~~~~~~~~~~~~~~~~~~ + \bar{A}^{}_{f^{}_1}
\bar{A}^{}_{f^{}_2} \left [ \tilde{g}^{}_+(t^{}_1) g^{}_-(t^{}_2)
+ C g^{}_-(t^{}_1) \tilde{g}^{}_+(t^{}_2) \right ] \right \} \; .
~~~~~
\end{eqnarray}
The calculation of the decay rate $R(f^{}_1, t^{}_1; f^{}_2,
t^{}_2)^{}_C \propto |A (f^{}_1, t^{}_1; f^{}_2, t^{}_2 )^{}_C|^2$
is straightforward but lengthy. Our result is
\begin{eqnarray}
R \left (f^{}_1, t^{}_1; f^{}_2, t^{}_2 \right )^{}_C & \propto &
|A^{}_{f^{}_1}|^2 |A^{}_{f^{}_2}|^2 e^{-\Gamma t^{}_+} \left [
\left (|\xi^{}_C|^2 + |\zeta^{}_C|^2 \right ) \cosh \left (y^{}_q
\Gamma t^{}_C \right ) \right . \nonumber \\
&& - \left (|\xi^{}_C|^2 - |\zeta^{}_C|^2 \right ) \cos \left
(x^{}_q \Gamma t^{}_C \right ) - 2{\rm Re}\left (\xi^*_C
\zeta^{}_C \right ) \sinh \left
(y^{}_q \Gamma t^{}_C \right ) \nonumber \\
&& \left . + 2{\rm Im}\left (\xi^*_C \zeta^{}_C \right ) \sin
\left (x^{}_q \Gamma t^{}_C \right ) + {\cal V} \left (f^{}_1,
t^{}_1; f^{}_2, t^{}_2 \right )^{}_C \right ] \; ,
~~~~~~~~~~~~~~~~~~~~~~~~
\end{eqnarray}
where $t^{}_C = t^{}_2 + C t^{}_1$ is defined, $\xi^{}_C$ and
$\zeta^{}_C$ have been given in Eq. (6), and ${\cal V} (f^{}_1,
t^{}_1; f^{}_2, t^{}_2)^{}_C$ denotes the $CPT$-violating term:
\begin{eqnarray}
{\cal V} \left (f^{}_1, t^{}_1; f^{}_2, t^{}_2 \right )^{}_-  & =
& - 2 {\rm Re} \left [ \left ( \xi^{}_- \zeta^*_- + \xi^*_-
\zeta^{}_+ \right ) \cos\theta \right ]
\cosh \left (y^{}_q \Gamma t^{}_- \right ) \nonumber \\
&& - 2 {\rm Re} \left [ \left ( \xi^{}_- \zeta^*_- - \xi^*_-
\zeta^{}_+ \right ) \cos\theta \right ]
\cos \left (x^{}_q \Gamma t^{}_- \right ) \nonumber \\
&& + 2 {\rm Re} \left [ \left ( |\xi^{}_-|^2 + \zeta^*_-
\zeta^{}_+ \right ) \cos\theta \right ]
\sinh \left (y^{}_q \Gamma t^{}_- \right ) \nonumber \\
&& - 2 {\rm Im} \left [ \left ( |\xi^{}_-|^2 - \zeta^*_-
\zeta^{}_+ \right ) \cos\theta \right ]
\sin \left (x^{}_q \Gamma t^{}_- \right ) \nonumber \\
&& +  \left | \xi^{}_- \left (\xi^{}_- + \zeta^{}_- \right )
\right | |\cos\theta| e^{+y^{}_q \Gamma t^{}_1} \cos \left (x^{}_q
\Gamma t^{}_2 - \Theta^{}_+ \right ) \nonumber \\
&& - \left | \xi^{}_- \left (\xi^{}_- - \zeta^{}_- \right ) \right
| |\cos\theta| e^{-y^{}_q \Gamma t^{}_1} \cos \left (x^{}_q \Gamma
t^{}_2 + \Theta^{}_- \right ) \nonumber \\
&& - \left | \xi^{}_- \left (\xi^{}_- - \zeta^{}_- \right ) \right
| |\cos\theta| e^{+y^{}_q \Gamma t^{}_2} \cos \left (x^{}_q \Gamma
t^{}_1 - \Theta^{}_- \right ) \nonumber \\
&& + \left | \xi^{}_- \left (\xi^{}_- + \zeta^{}_- \right ) \right
| |\cos\theta| e^{-y^{}_q \Gamma t^{}_2} \cos \left (x^{}_q \Gamma
t^{}_1 + \Theta^{}_+ \right ) \; ~~~~~~~~~~~~~~~~~~~~~~~
\end{eqnarray}
with
\begin{equation}
\tan\Theta^{}_{\pm} \; =\; \frac{\displaystyle {\rm Im} \left [
\xi^{}_- \left (\xi^*_- \pm \zeta^*_- \right ) \cos\theta \right
]}{\displaystyle {\rm Re} \left [ \xi^{}_- \left (\xi^*_- \pm
\zeta^*_- \right ) \cos\theta \right ]} \; ,
\end{equation}
or
\begin{eqnarray}
{\cal V} \left (f^{}_1, t^{}_1; f^{}_2, t^{}_2 \right )^{}_+  & =
& + 2 {\rm Re} \left ( \xi^{}_- \zeta^*_+ \cos\theta \right )
\cosh \left (y^{}_q \Gamma t^{}_+ \right ) \nonumber \\
&& + 2 {\rm Re} \left ( \xi^{}_- \zeta^*_+ \cos\theta \right )
\cos \left (x^{}_q \Gamma t^{}_+ \right ) \nonumber \\
&& - 2 {\rm Re} \left ( \xi^{}_- \xi^*_+ \cos\theta \right )
\sinh \left (y^{}_q \Gamma t^{}_+ \right ) \nonumber \\
&& + 2 {\rm Im} \left ( \xi^{}_- \xi^*_+ \cos\theta \right )
\sin \left (x^{}_q \Gamma t^{}_+ \right ) \nonumber \\
&& +  \left | \left (\xi^{}_+ - \zeta^{}_+ \right ) \left
(\xi^{}_- + \zeta^{}_- \right ) \right | |\cos\theta| e^{+y^{}_q
\Gamma t^{}_1} \cos \left (x^{}_q \Gamma
t^{}_2 + \Theta^{}_{-+} \right ) \nonumber \\
&& - \left | \left (\xi^{}_+ + \zeta^{}_+ \right ) \left (\xi^{}_-
- \zeta^{}_- \right ) \right | |\cos\theta| e^{-y^{}_q \Gamma
t^{}_1} \cos \left (x^{}_q \Gamma
t^{}_2 - \Theta^{}_{+-} \right ) \nonumber \\
&& + \left | \left (\xi^{}_+ - \zeta^{}_+ \right ) \left (\xi^{}_-
- \zeta^{}_- \right ) \right | |\cos\theta| e^{+y^{}_q \Gamma
t^{}_2} \cos \left (x^{}_q \Gamma
t^{}_1 + \Theta^{}_{--} \right ) \nonumber \\
&& - \left | \left (\xi^{}_+ + \zeta^{}_+ \right ) \left (\xi^{}_-
+ \zeta^{}_- \right ) \right | |\cos\theta| e^{-y^{}_q \Gamma
t^{}_2} \cos \left (x^{}_q \Gamma t^{}_1 - \Theta^{}_{++} \right )
\; ~~~~~~~~~~~~
\end{eqnarray}
with
\begin{equation}
\tan\Theta^{}_{\pm\pm} \; =\; \frac{\displaystyle {\rm Im} \left [
\left ( \xi^*_+ \pm \zeta^*_+ \right ) \left (\xi^{}_- \pm
\zeta^{}_- \right ) \cos\theta \right ]}{\displaystyle {\rm Re}
\left [ \left (\xi^*_+ \pm \zeta^*_+ \right ) \left (\xi^{}_- \pm
\zeta^{}_- \right ) \cos\theta \right ]} \; .
\end{equation}
We observe that the $CPT$-violating term has a more complicated
time-dependent behavior than the $CPT$-conserving term. For
completeness, the time-integrated form of $R(f^{}_1, t^{}_1;
f^{}_2, t^{}_2)^{}_C$ is given by
\begin{eqnarray}
R\left (f^{}_1, f^{}_2 \right )^{}_C & \propto & |A^{}_{f^{}_1}|^2
|A^{}_{f^{}_2}|^2 \left [ \frac{1 + Cy^2_q}{\left (1 - y^2_q
\right )^2} \left (|\xi^{}_C|^2 + |\zeta^{}_C|^2 \right ) -
\frac{1 - C x^2_q}{\left (1 + x^2_q \right )^2} \left
(|\xi^{}_C|^2 - |\zeta^{}_C|^2 \right ) \right . \nonumber \\
&& \left . - \frac{ 2 \left (1 + C \right ) y^{}_q}{\left (1 -
y^2_q \right )^2} {\rm Re} \left (\xi^*_C \zeta^{}_C \right ) +
\frac{2 \left (1 + C \right ) x^{}_q}{\left (1 + x^2_q \right )^2}
{\rm Im} \left (\xi^*_C \zeta^{}_C \right ) + 2 {\cal V} \left
(f^{}_1, f^{}_2 \right )^{}_C \right ] \; , ~~~~~
\end{eqnarray}
where the $CPT$-violating term reads
\begin{eqnarray}
{\cal V} \left (f^{}_1, f^{}_2 \right )^{}_- & = & - \frac{1}{1 -
y^2_q} {\rm Re} \left [ \left (\xi^{}_- \zeta^*_- + \xi^*_-
\zeta^{}_+ \right ) \cos\theta \right ] \nonumber \\
&& - \frac{1}{1 + x^2_q} {\rm Re} \left [ \left (\xi^{}_-
\zeta^*_- - \xi^*_- \zeta^{}_+ \right ) \cos\theta \right ] \nonumber \\
&& + \frac{2}{\left (1 + x^2_q \right ) \left (1 - y^2_q \right )}
\left [ {\rm Re} \left ( \xi^{}_- \zeta^*_- \cos\theta \right ) +
x^{}_q y^{}_q {\rm Im} \left (\xi^{}_- \zeta^*_- \cos\theta \right
) \right ] \; , ~~~~~~~~~~~
\end{eqnarray}
or
\begin{eqnarray}
{\cal V} \left (f^{}_1, f^{}_2 \right )^{}_+ & = & + \left [
\frac{1 + y^2_q}{\left (1 - y^2_q \right )^2} + \frac{1 -
x^2_q}{\left (1 + x^2_q \right )^2} \right ] {\rm Re} \left (
\xi^{}_- \zeta^*_+ \cos\theta \right ) \nonumber \\
&& - \frac{2 y^{}_q}{\left (1 - y^2_q \right )^2} \cdot
\frac{x^2_q + y^2_q}{1 + x^2_q} {\rm Re} \left
(\xi^{}_- \xi^*_+ \cos\theta \right ) \nonumber \\
&& - \frac{2 x^{}_q}{\left (1 + x^2_q \right )^2} \cdot
\frac{x^2_q + y^2_q}{1 - y^2_q} {\rm Im} \left
(\xi^{}_- \xi^*_+ \cos\theta \right ) \nonumber \\
&& - \frac{2}{\left (1 + x^2_q \right ) \left (1 - y^2_q \right )}
\left [ {\rm Re} \left ( \xi^{}_- \zeta^*_+ \cos\theta \right ) +
x^{}_q y^{}_q {\rm Im} \left (\xi^{}_- \zeta^*_+ \cos\theta \right
) \right ] \; . ~~~~~~~~~~~~
\end{eqnarray}
When $CPT$ is a good symmetry ($\cos\theta =0$), Eqs. (9) and (14)
are in agreement with the formulas obtained in Refs.
\cite{Xing97,XingP}.

We stress that Eqs. (9)--(16) are new results and may serve as the
master equations for the study of $CPT$ violation in coherent
$(B^0_q\bar{B}^0_q)^{}_C$ decays. They are also valid for other
correlated meson-antimeson systems with $CPT$ violation, in
particular useful to describe coherent $(D^0\bar{D}^0)^{}_C$
decays at the $\psi (3770)$ and $\psi (4140)$ resonances.

\section{Example: opposite-sign dilepton events}

To be specific, we consider a particularly simple and interesting
possibility of testing $CPT$ invariance in $B^0_q$-$\bar{B}^0_q$
mixing: the opposite-sign dilepton events from coherent
$(B^0_q\bar{B}^0_q)^{}_C$ decays. One may take $f^{}_1 = l^+_1
X^-_1$ and $f^{}_2 = l^-_2 X^+_2$ with either $l^{}_1 = l^{}_2$
(e.g., $l^{}_1 = l^{}_2 = \mu$) or $l^{}_1 \neq l^{}_2$ (e.g.,
$l^{}_1 = e$ and $l^{}_2 =\mu$). The amplitude of each
semileptonic decay mode can be parameterized in analogy with Eq.
(5). Since the $\Delta B = -\Delta Q$ transitions must be strongly
suppressed, it is reasonable to take $|\sigma^{}_{l^{}_i}| \ll 1$
(for $i =1$ or $2$) in our calculations.

We first look at the $C=-1$ case. By simplifying Eqs. (9), (10)
and (11), we obtain the time-dependent rates of
$(B^0_q\bar{B}^0_q)^{}_- \rightarrow (l^\pm_1 X^\mp_1)^{}_{t^{}_1}
(l^\mp_2 X^\pm_2)^{}_{t^{}_2}$ decays as follows:
\begin{eqnarray}
R\left (l^+_1 X^-_1, t^{}_1; l^-_2 X^+_2, t^{}_2 \right )^{}_- &
\propto & |A^{}_{l^{}_1}|^2 |A^{}_{l^{}_2}|^2  e^{-\Gamma t^{}_+}
\left [ \cosh (y^{}_q \Gamma t^{}_- ) + \cos (x^{}_q \Gamma t^{}_-
) \right . \nonumber \\
&& \left . +2 {\rm Re}\Omega \sinh (y^{}_q \Gamma t^{}_- ) + 2
{\rm Im}\Omega \sin (x^{}_q \Gamma t^{}_- ) \right ] \; ,
\nonumber \\
R\left (l^-_1 X^+_1, t^{}_1; l^+_2 X^-_2, t^{}_2 \right )^{}_- &
\propto & |A^{}_{l^{}_1}|^2 |A^{}_{l^{}_2}|^2  e^{-\Gamma t^{}_+}
\left [ \cosh (y^{}_q \Gamma t^{}_- ) + \cos (x^{}_q \Gamma t^{}_-
) \right . \nonumber \\
&& \left . -2 {\rm Re}\overline{\Omega} \sinh (y^{}_q \Gamma
t^{}_- ) - 2 {\rm Im}\overline{\Omega} \sin (x^{}_q \Gamma t^{}_-
) \right ] \; ,
\end{eqnarray}
where $t^{}_{\pm} = t^{}_2 \pm t^{}_1$, and
\begin{equation}
\Omega \; =\; \cos\theta + \sigma^{}_{l^{}_1} e^{+i\phi} -
\sigma^*_{l^{}_2} e^{-i\phi} \; , ~~~~~~~~~~~~ \overline{\Omega}
\; =\; \cos\theta - \sigma^*_{l^{}_1} e^{-i\phi} +
\sigma^{}_{l^{}_2} e^{+i\phi} \;
\end{equation}
have been defined by keeping the leading terms of $CPT$ violation
and $\Delta B =-\Delta Q$ effects. Note that $e^{i\phi} =
(V^*_{tb} V^{}_{ts})/(V^{}_{tb} V^*_{ts}) \approx 1 + 2 i
\lambda^2\eta$ holds for $B^0_s$-$\bar{B}^0_s$ mixing described by
the box diagrams in the standard model, where $\lambda \approx
0.22$ and $\eta \approx 0.34$ are the Wolfenstein parameters
\cite{PDG06}. If $e^{\pm i \phi} \approx 1$ is taken in the
leading-order approximation, Eq. (18) can be simplified to $\Omega
\approx \cos\theta + (\sigma^{}_{l^{}_1} - \sigma^*_{l^{}_2})$ and
$\overline{\Omega} \approx \cos\theta - (\sigma^*_{l^{}_1} -
\sigma^{}_{l^{}_2})$. If $l^{}_1 = l^{}_2$ is further taken, then
we have $\Omega \approx \overline{\Omega} \approx \cos\theta + 2 i
{\rm Im}(\sigma^{}_{l^{}_1})$. It is remarkable that the same
simplification cannot be made for the $B^0_d$-$\bar{B}^0_d$ mixing
system, where $e^{i\phi} = (V^*_{tb} V^{}_{td})/(V^{}_{tb}
V^*_{td}) \approx e^{-2i\beta}$ with $\beta \approx 22^\circ$
being one of the inner angles of the Cabibbo-Kobayashi-Maskawa
unitarity triangle in the standard phase convention \cite{PDG06}.
Of course, $e^{i\phi}$ might deviate from the standard-model
expectation if $B^0_q$-$\bar{B}^0_q$ mixing (for $q=d$ or $s$)
involves a kind of new physics. The latter has also been included
into the parameters $\Omega$ and $\overline{\Omega}$. Thus these
two parameters serve for an effective description of possible new
physics ($CPT$ violation, $\Delta B = -\Delta Q$ transitions and
new $\Delta B =2$ effects) in $B^0_q$-$\bar{B}^0_q$ mixing.

Eqs. (17) and (18) clearly show that the $\Delta B =-\Delta Q$
parameters have the same time-dependent behavior as the
$CPT$-violating parameter $\cos\theta$ in the opposite-sign
dilepton events of the correlated $B^0_q\bar{B}^0_q$ state with
$C=-1$. Hence it is in general impossible to distinguish between
the effect of $CPT$ violation and that of $\Delta B =-\Delta Q$
transitions in this kind of events, unless one of them is
remarkably smaller than the other. If the decays of the correlated
$B^0_q\bar{B}^0_q$ state with $C=+1$ are taken into account,
however, it is in principle possible to cleanly extract the
$CPT$-violating parameter \cite{Grimus}. To illustrate this point
in a transparent way, we simplify Eqs. (9), (12) and (13) to
obtain the time-dependent rates of $(B^0_q\bar{B}^0_q)^{}_+
\rightarrow (l^\pm_1 X^\mp_1)^{}_{t^{}_1} (l^\mp_2
X^\pm_2)^{}_{t^{}_2}$ decays. The result is
\begin{eqnarray}
R\left (l^+_1 X^-_1, t^{}_1; l^-_2 X^+_2, t^{}_2 \right )^{}_+ &
\propto & |A^{}_{l^{}_1}|^2 |A^{}_{l^{}_2}|^2  e^{-\Gamma t^{}_+}
\left [ \cosh (y^{}_q \Gamma t^{}_+ ) + \cos (x^{}_q \Gamma t^{}_+
) \right . \nonumber \\
&& \left . -2 {\rm Re}\Omega^\prime \sinh (y^{}_q \Gamma t^{}_+ )
- 2 {\rm Im}\Omega^\prime \sin (x^{}_q \Gamma t^{}_+ ) + 2 \Delta
(t^{}_1, t^{}_2) \right ] , ~~~~~~ \nonumber \\
R\left (l^-_1 X^+_1, t^{}_1; l^+_2 X^-_2, t^{}_2 \right )^{}_+ &
\propto & |A^{}_{l^{}_1}|^2 |A^{}_{l^{}_2}|^2  e^{-\Gamma t^{}_+}
\left [ \cosh (y^{}_q \Gamma t^{}_+ ) + \cos (x^{}_q \Gamma t^{}_+
) \right . \nonumber \\
&& \left . -2 {\rm Re}\overline{\Omega}^\prime \sinh (y^{}_q
\Gamma t^{}_+ ) - 2 {\rm Im}\overline{\Omega}^\prime \sin (x^{}_q
\Gamma t^{}_+ ) - 2 \Delta (t^{}_1, t^{}_2) \right ] , ~~~~~~
\end{eqnarray}
where $\Omega^\prime = \sigma^{}_{l^{}_1} e^{+i\phi} +
\sigma^*_{l^{}_2} e^{-i\phi}$ and $\overline{\Omega}^\prime =
\sigma^*_{l^{}_1} e^{-i\phi} + \sigma^{}_{l^{}_2} e^{+i\phi}$ do
not contain the $CPT$-violating effect, but $\Delta (t^{}_1,
t^{}_2)$ is purely a $CPT$-violating term:
\begin{eqnarray}
\Delta (t^{}_1, t^{}_2) & = & + \left [ \cos (x^{}_q \Gamma t^{}_1
) \sinh (y^{}_q \Gamma t^{}_2 ) - \sinh (y^{}_q \Gamma t^{}_1 )
\cos (x^{}_q \Gamma t^{}_2 ) \right ] {\rm Re} (\cos\theta)
\nonumber \\
&& - \left [ \sin (x^{}_q \Gamma t^{}_1 ) \cosh (y^{}_q \Gamma
t^{}_2 ) - \cosh (y^{}_q \Gamma t^{}_1 ) \sin (x^{}_q \Gamma
t^{}_2 ) \right ] {\rm Im} (\cos\theta) \; . ~~~~~~~~~~~~~~~~~
\end{eqnarray}
One can easily see that $\Delta (t^{}_2, t^{}_1) = - \Delta
(t^{}_1, t^{}_2)$ holds, but the $CPT$-conserving terms in Eq.
(19) do not change with the exchange of $t^{}_1$ and $t^{}_2$.
This interesting feature implies that $\Delta (t^{}_1, t^{}_2)$
can in principle be extracted from the rate differences
\begin{eqnarray}
R\left (l^+_1 X^-_1, t^{}_1; l^-_2 X^+_2, t^{}_2 \right )^{}_+ -
R\left (l^+_1 X^-_1, t^{}_2; l^-_2 X^+_2, t^{}_1 \right )^{}_+ &
\propto & 4 |A^{}_{l^{}_1}|^2 |A^{}_{l^{}_2}|^2  e^{-\Gamma
t^{}_+} \Delta (t^{}_1, t^{}_2) \; , \nonumber \\
R\left (l^-_1 X^+_1, t^{}_2; l^+_2 X^-_2, t^{}_1 \right )^{}_+ -
R\left (l^-_1 X^+_1, t^{}_1; l^+_2 X^-_2, t^{}_2 \right )^{}_+ &
\propto & 4 |A^{}_{l^{}_1}|^2 |A^{}_{l^{}_2}|^2  e^{-\Gamma
t^{}_+} \Delta (t^{}_1, t^{}_2) \; . ~~~~~~~
\end{eqnarray}
As ${\rm Im}\phi \approx 0$ is a good approximation for
$B^0_q$-$\bar{B}^0_q$ mixing in the standard model, we have ${\rm
Re} \overline{\Omega}^\prime \approx {\rm Re} \Omega^\prime$ and
${\rm Im} \overline{\Omega}^\prime \approx -{\rm Im}
\Omega^\prime$. In this case, ${\rm Im}\Omega^\prime$ can in
principle be extracted from the rate differences
\begin{eqnarray}
R\left (l^-_1 X^+_1, t^{}_1; l^+_2 X^-_2, t^{}_2 \right )^{}_+ -
R\left (l^+_1 X^-_1, t^{}_2; l^-_2 X^+_2, t^{}_1 \right )^{}_+ &
\propto & 4 |A^{}_{l^{}_1}|^2 |A^{}_{l^{}_2}|^2  e^{-\Gamma
t^{}_+} {\rm Im}\Omega^\prime \; ,
\nonumber \\
R\left (l^-_1 X^+_1, t^{}_2; l^+_2 X^-_2, t^{}_1 \right )^{}_+ -
R\left (l^+_1 X^-_1, t^{}_1; l^-_2 X^+_2, t^{}_2 \right )^{}_+ &
\propto & 4 |A^{}_{l^{}_1}|^2 |A^{}_{l^{}_2}|^2  e^{-\Gamma
t^{}_+} {\rm Im}\Omega^\prime \; . ~~~~~~~~~~~ \;
\end{eqnarray}
For $B^0_s$-$\bar{B}^0_s$ mixing with $e^{\pm i\phi} \approx 1$,
${\rm Im} \Omega^\prime \approx {\rm Im} \sigma^{}_{l^{}_1} - {\rm
Im} \sigma^{}_{l^{}_2}$ holds; and for $B^0_d$-$\bar{B}^0_d$
mixing with $e^{\pm i\phi} \approx e^{\mp 2i\beta}$, we obtain
${\rm Im} \Omega^\prime \approx ({\rm Re}\sigma^{}_{l^{}_2} - {\rm
Re}\sigma^{}_{l^{}_1}) \sin 2\beta - ({\rm Im}\sigma^{}_{l^{}_2} -
{\rm Im}\sigma^{}_{l^{}_1}) \cos 2\beta$. Thus the dilepton events
of coherent $(B^0_q\bar{B}^0_q)^{}_+$ decays are very useful to
probe possible $CPT$ violation and $\Delta B = -\Delta Q$ effects.

If the forthcoming super-$B$ factory is also an asymmetric
$e^+e^-$ collider as the present KEK and SLAC $B$-meson factories,
it will be easier to measure the proper-time difference $t^{}_- =
(t^{}_2 - t^{}_1)$ of a dilepton event. A measurement of the
$t^{}_+ = (t^{}_2 + t^{}_1)$ distribution might be difficult in
either linacs or storage rings, unless the bunch lengths are much
shorter than the decay lengths \cite{Plus}. Hence we may calculate
the $t^{}_-$ distribution of the dilepton events by integrating
$R(l^\pm_1 X^\mp_1, t^{}_1; l^\mp_2 X^\pm_2, t^{}_2)^{}_C$ over
$t^{}_+$. For simplicity, here we assume $\Delta B = \Delta Q$ to
be a perfect rule and use $t$ to denote $t^{}_-$. We take $t>0$ by
convention. Our results are
\begin{eqnarray}
R\left (l^\pm_1 X^\mp_1, l^\mp_2 X^\pm_2, t \right )^{}_- &
\propto & |A^{}_{l^{}_1}|^2 |A^{}_{l^{}_2}|^2  e^{-\Gamma t} \left
[ \cosh (y^{}_q \Gamma t) + \cos (x^{}_q \Gamma t)
\right . \nonumber \\
&& \left . \pm 2 {\rm Re}(\cos\theta) \sinh (y^{}_q \Gamma t) \pm
2 {\rm Im}(\cos\theta) \sin (x^{}_q \Gamma t) \right ] \; ,
~~~~~~~~~~~~~~~~\;
\end{eqnarray}
and
\begin{eqnarray}
R\left (l^\pm_1 X^\mp_1, l^\mp_2 X^\pm_2, t \right )^{}_+ &
\propto & |A^{}_{l^{}_1}|^2 |A^{}_{l^{}_2}|^2  e^{-\Gamma t} \left
[ \frac{\cosh (y^{}_q \Gamma t + \varphi^{}_y)}{\sqrt{1 - y^2_q}}
+ \frac{\cos (x^{}_q \Gamma t + \varphi^{}_x)}{\sqrt{1 +
x^2_q}} \right . \nonumber \\
&& \pm \frac{2 |\cos\theta| \left [ \cos (\Theta + \omega^{}_-)
e^{+y^{}_q \Gamma t} - \cos (\Theta + \omega^{}_- + x^{}_q \Gamma
t) \right ]}{\sqrt{x^2_q + (2 - y^{}_q)^2}} \nonumber \\
&& \left . \mp \frac{2 |\cos\theta| \left [ \cos (\Theta -
\omega^{}_+) e^{-y^{}_q \Gamma t} - \cos (\Theta - \omega^{}_+ -
x^{}_q \Gamma t) \right ]}{\sqrt{x^2_q + (2 + y^{}_q)^2}} \right ]
\; , ~~~~~~
\end{eqnarray}
where the parameters $\varphi^{}_x$, $\varphi^{}_y$,
$\omega^{}_\pm$ and $\Theta$ are defined through $\tan\varphi^{}_x
= x^{}_q$, $\tanh\varphi^{}_y = y^{}_q$, $\tan\omega^{}_\pm =
x^{}_q/(2\pm y^{}_q)$ and $\tan\Theta = {\rm Im}(\cos\theta)/{\rm
Re}(\cos\theta)$, respectively. Taking $x^{}_s \sim 27$ and
$y^{}_s \sim 0.07$ for example, we have $\varphi^{}_x \sim 1.53$,
$\varphi^{}_y \sim 0.07$, $\omega^{}_+ \sim 1.49$ and $\omega^{}_-
\sim 1.50$; while taking $x^{}_d \sim 0.78$ and $y^{}_d \sim
0.002$ for example, we have $\varphi^{}_x \sim 0.66$,
$\varphi^{}_y \sim 0.002$ and $\omega^{}_+ \approx \omega^{}_-
\sim 0.37$. Eqs. (23) and (24) show that both ${\rm
Re}(\cos\theta)$ and ${\rm Im}(\cos\theta)$ can in principle be
determined or constrained by measuring the decay rates $R(l^\pm_1
X^\mp_1, t^{}_1; l^\mp_2 X^\pm_2, t^{}_2)^{}_C$, provided the
$\Delta B = -\Delta Q$ transitions and other new-physics effects
are negligibly small. These formulas are also applicable for the
$D^0$-$\bar{D}^0$ mixing system.

\section{Summary}

Keeping with the great experimental interest in testing discrete
symmetries and conservation laws at the present and future
$B$-meson factories, we have reformulated the time-dependent and
time-integrated rates of coherent $(B^0_q\bar{B}^0_q)^{}_C$ decays
($q=d$ or $s$) by assuming small $CPT$ violation in $B^0_q$-$
\bar{B}^0_q$ oscillation. Our results are new and generic, and
thus they can serve as the master formulas for the analysis of
possible $CPT$-violating effects in both the $B^0_d$-$\bar{B}^0_d$
mixing system at the $\Upsilon (4S)$ resonance and the
$B^0_s$-$\bar{B}^0_s$ mixing system at the $\Upsilon (5S)$
resonance. Taking the opposite-sign dilepton events for example,
we have shown that it is possible to separately determine or
constrain the parameters of $CPT$ violation and $\Delta B =
-\Delta Q$ transitions by measuring their time distributions in
the $C=+1$ case. In the $C=-1$ case, however, the $CPT$-violating
and $\Delta B = -\Delta Q$ effects have the same time-dependent
behavior and are in general indistinguishable.

We expect that a stringent test of $CPT$ symmetry and the $\Delta
B =\Delta Q$ rule will finally be realized at a super-$B$ factory
with the luminosity ${\cal L} \sim {\rm a ~ few} \times
10^{36}{\rm cm}^{-2}{\rm s}^{-1}$, where other kinds of new
physics may also be explored. The prospect of such ambitious
experiments is by no means dim, indeed.

Finally let us mention the evidence for $D^0$-$\bar{D}^0$ mixing
achieved in the BaBar \cite{BB1} and Belle \cite{BE} experiments.
It turns out that the mixing parameter $y^{}_D$ is at the one
percent level and $|x^{}_D| < |y^{}_D|$ is expected to hold
\cite{Nir}. Therefore it seems possible to test $CP$, $T$ and
$CPT$ symmetries in the charm system in the (far) future. As we
have emphasized before, the master formulas obtained in this paper
can all be used to describe coherent $D^0\bar{D}^0$ decays with
small $CPT$ violation and (or) small $\Delta C = -\Delta Q$
effects.

\vspace{0.5cm}

This work was supported in part by the National Natural Science
Foundation of China.

\end{document}